# Photometry of the dwarf nova AW Sagittae during the 2006 November superoutburst

Jeremy Shears, Roger Pickard, Tom Krajci, & Gary Poyner

**Abstract**

During 2006 November an outburst of the dwarf nova AW Sge was observed using CCD photometry. This revealed 0.25 magnitude superhumps confirming it to be a superoutburst, possibly only the second confirmed such outburst of this star. The superhumps were observed for 4 days and had a stable period $P_{sh}$ = 0.0745(2) d, a value which is consistent with $P_{sh}$ measured during the 2000 superoutburst.

**Introduction**

Dwarf novae are a class of cataclysmic variable star in which a white dwarf primary accretes material from a secondary star via Roche lobe overflow. The secondary is usually a late-type main-sequence star. In the absence of a significant white dwarf magnetic field, material from the secondary is processed through an accretion disc before settling on the surface of the white dwarf. As material builds up in the disc, a thermal instability is triggered that drives the disc into a hotter, brighter state causing an outburst in which the star brightens by several magnitudes [1]. Dwarf novae of the SU UMa family occasionally exhibit superoutbursts which last several times longer than normal outbursts and may be up to a magnitude brighter. During a superoutburst the light curve of a SU UMa star is characterised by superhumps. These are modulations in the light curve which are a few percent longer than the orbital period. They are thought to arise from the interaction of the secondary star orbit with a slowly precessing eccentric accretion disc. The eccentricity of the disc arises because a 3:1 resonance occurs between the secondary star orbit and the motion of matter in the outer accretion disc. For a more detailed review of SU UMa stars and superhumps, the reader is directed to references 2 and 3.

**History of AW Sge**

AW Sge was first reported by M. and G. Wolf during their photographic search for variable stars in the vicinity of gamma Sge conducted in the opening years of the twentieth century [4]. They detected it on two occasions, in 1901 and 1905, at magnitude 13 and 15 respectively, whereas on a further 5 occasions between 1900 and late 1905 it was not visible. The star was subsequently included in Vogt and Bateson's atlas of possible dwarf novae by virtue of its blue colour [5].

| Date (UT) | Magnitude at discovery | Observer |
|---|---|---|
| 1996 Sep 17.900 | 15.0C | J. Pietz [6] |
| 2000 Jul 11.565 | 14.0v | R. Stubbings [7] |
| 2002 Oct 31.036 | 14.4v | M. Simonsen |
| 2004 May 22.444 | 14.8CR | P. Schmeer |
| 2006 Nov 16.764 | 14.8C | J. Shears |

**Table 1:** Recent reported outbursts of AW Sge
Unless specified otherwise, data are from the AAVSO International Database
C= unfiltered CCD, CR = unfiltered CCD, calibrated with R (red) comparison star sequence, v= visual

The first confirmed outburst of AW Sge in modern times was detected by J. Pietz in 1996 [6] and to date a total of 5 outbursts have been reported, including the one discussed in this paper (Table 1). The shortest time between outbursts is 569 days and the longest is 1393 days, with a median outburst interval of 890 days. However, it is entirely possible that other outbursts have been missed.

To investigate this further, we examined the AAVSO International Database, which contains almost five thousand observations of AW Sge since 1981 August, for other outbursts. This was not straightforward as we found that several observers always recorded AW Sge as impossibly bright at times when other observers reported no outburst: some had it routinely ~ $12^{th}$ magnitude and some at ~14.8-15.1. We suspect that another star has been mistaken for the variable in what is a rich field. Hence we decided to omit observations by these observers, especially as there was no independent confirmation by other observers. The remaining observations were either negative observations (mainly visual), or very faint CCD measurements of the star at quiescence (mag 18 to 19). Thus we can find no evidence of outbursts other than those in Table 1. However, seasonal gaps of 2 to 3 months are common in the observational record presumably due to the field being poorly located near the sun or else inconveniently located in the pre-dawn sky, a time when fewer observers are active.

Photometry by Henden in 1999 June showed AW Sge at 19.3V, which we take as its magnitude at quiescence [8]. Based on the brightness of the 2000 outburst at discovery, this suggests an outburst amplitude of at least 5.3 magnitudes.

Time resolved photometry during the 2000 outburst by G. Masi and M.A. Tosti revealed the presence of superhumps confirming this was a superoutburst and thus identifying AW Sge as a member of the SU UMa family [9]. The superhump period, $P_{sh}$, was quoted as 0.0745 d in a paper by Kato *et al* which cites Masi as the source [3]. During this outburst AW Sge's position was determined by Masi as RA 19h 58m 37.11s, Dec +16° 41' 28.8" (J2000).

According to observations in the AAVSO International Database, the 2000 superoutburst lasted at least 8 days. By contrast, the outbursts in 1996, 2002 and 2004 lasted 4 days or less and are therefore more likely to have been normal outbursts. We note that ~4 % of observations AAVSO International Database were separated by more than 4 days, so again outbursts could have been missed.

In an attempt to encourage monitoring of the star for outbursts, AW Sge was added to the BAA Variable Star Section's (VSS) Recurrent Objects Programme in 1994 [10]. This programme was set up as a joint project between the VSS and *The Astronomer* magazine specifically to monitor poorly studied eruptive stars of various types where outbursts occur at intervals of greater than 1 year. Observing charts and sequences for AW Sge are available from the AAVSO [11].

**Detection and course of the outburst**

The outburst reported here was first detected by JS on 2006 Nov 16.764 at 14.8C [12], rising to 14.6C by Nov 16.819 (Figure 1) and was independently detected later the same night by GP on Nov 16.846 at 14.4v [13]. This was significantly fainter than the 2000 outburst, which could indicate that the outburst was already well advanced and beginning to decline, however the AAVSO International Database records a negative observation by Eddie Muyllaert at <14.0v on Nov 15.796, only ~24 hours before the outburst was detected.

Figure 2 shows the overall light curve of the outburst covering the 8 days for which positive observations exist. During the first 4 days following detection, the star faded at an average rate of 0.12 mag/d. This is typical of a dwarf nova in the initial stages of decline. There then followed a more rapid decline at 0.38 mag/d, which is again typical of the final stages of a dwarf nova outburst.

**Detection of superhumps**

Six time-series photometry runs were conducted during the outburst yielding more than one thousand individual CCD images and totalling 17.5 hours of data. Table 2 summarises the instrumentation used and Table 3 contains a log of the time-series runs. In all cases raw images were flat-fielded and dark-subtracted, before being analysed using commercially available photometry software: JS and RP employed *AIP4WIN* version 1[14] with GSC1616-0855 (13.223V) as the comparison star, whereas TK employed *AIP4WIN* version 2 [15] with a comparison star ensemble of GSC1616-0277 and 1616-0687 (mag 11.977V and 12.095V). Unfortunately each run was necessarily rather short due to the unfavourable position of this object in the November evening sky.

The first time-series photometry data obtained on Nov 16 (Figure 3) shows a 0.25 mag hump-like feature in the light curve. However, the run was curtailed due to the field being obscured by local obstructions and it was not long enough to definitively characterise this feature as a superhump. Nevertheless, 0.25 to 0.3 mag superhumps were detected with certainty on Nov 17, 18, and 19; by Nov 20 they were slightly smaller, having an amplitude of 0.2 mag (Figures 4 to 6).

The detection of superhumps confirms this was a superoutburst, the first seen since 2000 July. The two superoutbursts were separated by more than 2300 days, however, as discussed above, it is entirely possible that other superoutbursts have been missed.

| Observer | Instrumentation |
|---|---|
| JS | 0.1-m fluorite refractor + Starlight Xpress SXV-M7 CCD |
| TK | 0.28-m SCT + SBIG ST-7E CCD |
| RP | 0.30-m SCT+ Starlight Xpress MX716 CCD + V filter |

**Table 2: Instrumentation used**

| Run number | Date (UT) | Start time JD-2454000 | Duration | No. of images | Mean mag. (C) | Observer |
|---|---|---|---|---|---|---|
| 1 | Nov 16 | 56.267 | 1.9 h | 79 | 14.81 | JS |
| 2 | Nov 17 | 57.237 | 3.8 h | 167 | 14.88 | RP |
| 3 | Nov 18 | 58.230 | 3.4 h | 153 | 15.01 | RP |
| 4 | Nov 19 | 58.531 | 3.4 h | 326 | 15.06 | TK |
| 5 | Nov 20 | 59.537 | 3.1 h | 214 | 15.25 | TK |
| 6 | Nov 20 | 60.291 | 1.9 h | 81 | 15.28 | RP |

**Table 3: Log of time-series observations**

**Measurement of the superhump period**

In order to study the superhump behaviour, we first extracted the times of each resolvable superhump maximum from the individual light curves according to the Kwee and van Woerden method [16] using the Peranso software [17]. Times of 9 superhump maxima were found and these were then used to assign superhump cycle numbers which best fitted the assumption of a constant superhump period. We found that the maxima fitted well a constant superhump period $P_{sh}$ = 0.0745(2) d, with a superhump maximum ephemeris JD 2454056.3241 + 0.0745(2) * E. The superhump cycle number, the measured times of superhump maximum and the O-C (Observed-Calculated) residuals relative to the above superhump maximum ephemeris are listed in Table 4 and a plot of residuals versus superhump cycle number is shown in Figure 7.

| Superhump cycle number | Time of maximum (JD) 2454000+ | O-C (cycles) |
|---|---|---|
| 0 | 56.3236 | -0.0067 |
| 13 | 57.2926 | +0.0013 |
| 14 | 57.3671 | -0.0242 |
| 26 | 58.2611 | -0.0027 |
| 27 | 58.3356 | -0.0107 |
| 30 | 58.5591 | +0.0134 |
| 31 | 58.6336 | -0.0080 |
| 44 | 59.6021 | +0.0282 |
| 54 | 60.3471 | -0.0295 |

**Table 4: Timing of superhump maxima**

To confirm our measurement of $P_{sh}$, we carried out a period analysis of all the data from the six time series runs using the ANOVA (Analysis of Variance) algorithm in Peranso, after subtracting the mean and linear trend from each of the light curves. This gave the power spectrum in Figure 8 which has its highest peak at a period of 0.0744(4) d, consistent with our earlier finding. The superhump period error estimate is derived using the Schwarzenberg-Czerny method [18]. Removing $P_{sh}$ from the power spectrum leaves only weak signals, none of which have any significant relationship to the superhump or orbital periods. A phase diagram of the data folded on this period is shown in Figure 9 which shows that the superhumps from each separate observing run become superimposed on each other. Hence this result also supports the assertion that the $P_{sh}$ remained remarkably stable during the 4 days over which the outburst was followed. However, our observations cover only the first part of the superoutburst, up to the point where a more rapid fade begins; in some SU UMa stars, period changes only occur late in the outburst [2].

We note that a superhump period of 0.0745(2) d (107.3 min) is consistent with the value reported by Kato *et al* based on observations made by Masi and Tosti during the only other known superoutburst in 2000 [9].

Stolz and Schoembs developed an empirical relationship between the orbital period $P_{orb}$ and the superhump period excess $\varepsilon = (P_{sh} - P_{orb}) / P_{orb}$ in dwarf novae [19]. Using their equation (6) and our value of $P_{sh}$ of 0.0745(2) d, it is possible to estimate that $\varepsilon \sim 0.030(8)$ and therefore $P_{orb} \sim 0.0723(7)$ d. However, radial velocity measurements or photometry at quiescence are required to measure $P_{orb}$ accurately.

**Future observations**

As discussed above, it is possible that outbursts of AW Sge have been missed because the normal outbursts are rather short (4 days or less) and because seasonal gaps of 2 to 3 months exist in the observational record. Hence, we encourage observers, whether visual or CCD-equipped, to monitor AW Sge for future outbursts which will help determine the true outburst frequency and the length of its supercycle. Observations at the beginning and the end of each observing season would be particularly valuable to minimise the effect of seasonal gaps.

**Acknowledgements**

The authors gratefully acknowledge the use of observations from the AAVSO International Database contributed by observers worldwide and the use of SIMBAD, operated through the Centre de Données Astronomiques (Strasbourg, France). We thank our two referees, Dr. Darren Baskill and Dr. Chris Lloyd, for their very constructive comments which have improved the paper. Finally, we are indebted to Dr. David Boyd for much advice during the execution of this project.

**Addresses:**
JS: "Pemberton", School Lane, Bunbury, Tarporley, Cheshire, CW6 9NR, UK
[bunburyobservatory@hotmail.com]
RP: 3 The Birches, Shobdon, Leominster, Herefordshire, HR6 9NG
[rdp@astronomy.freeserve.co.uk]
TK: CBA New Mexico, PO Box 1351 Cloudcroft, New Mexico 88317, USA
[tom_krajci@tularosa.net]
GP: 67 Ellerton Road, Kingstanding, Birmingham, B44 0QE, UK
[garypoyner@blueyonder.co.uk]

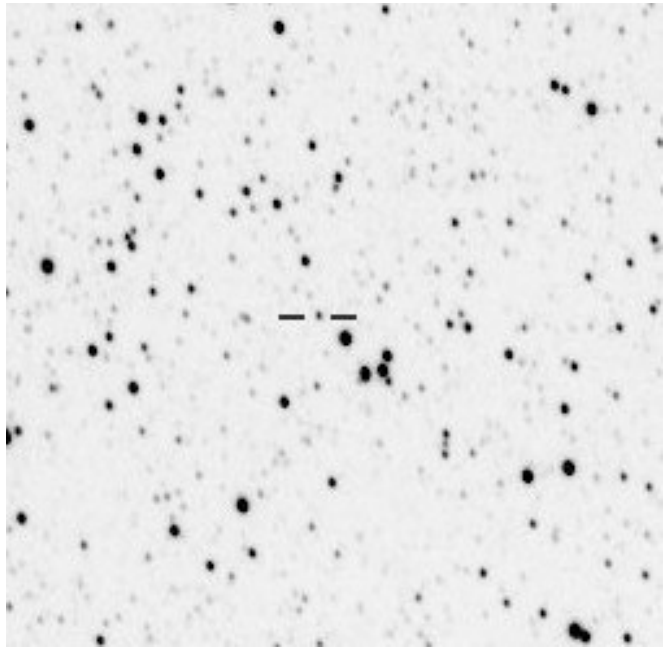

**Figure 1: AW Sge in outburst at 14.8C on 2006 Nov 16.764**
8' x 8' with south at top

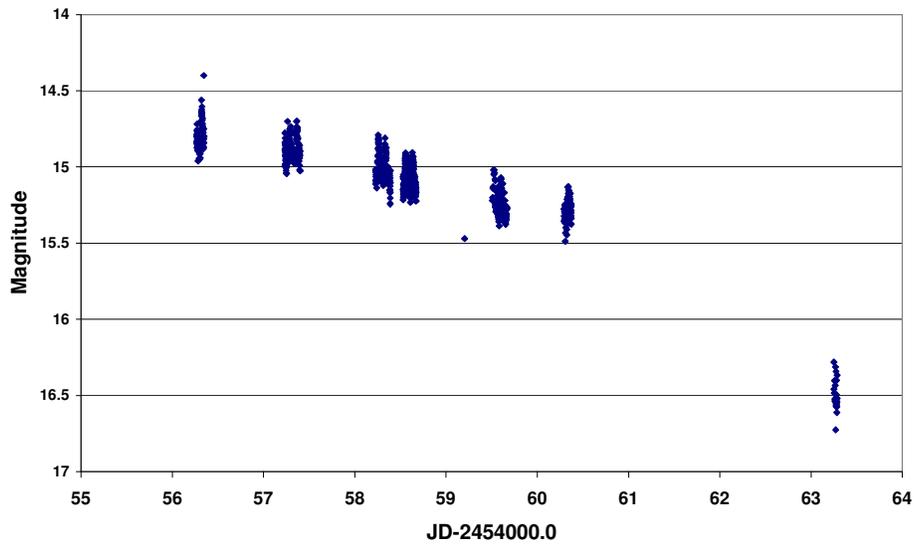

**Figure 2: Light curve of the 2006 November outburst**
Unfiltered or V (Visual) filtered CCD measurements; data is from the authors, plus the AAVSO International Database.

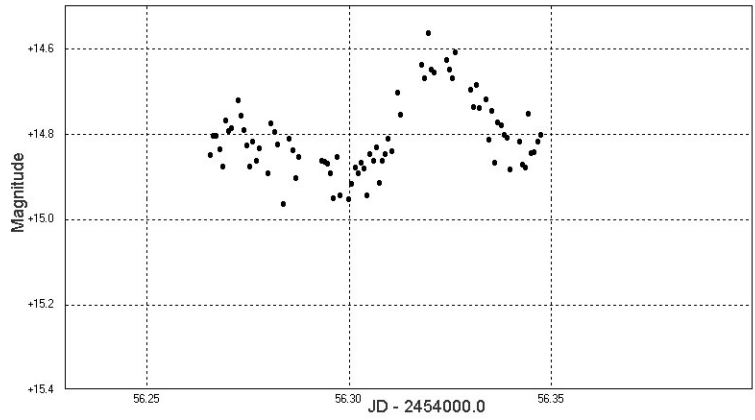
Figure 3: Time-series data from Nov 16 (J. Shears)

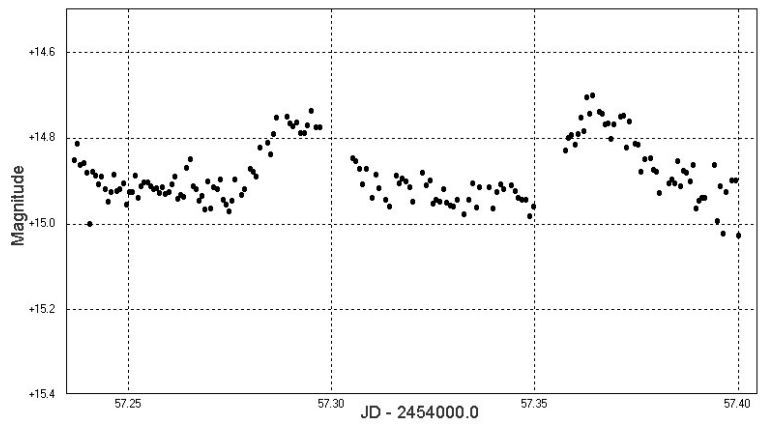
Figure 4: Time-series data from Nov 17 (R. Pickard)

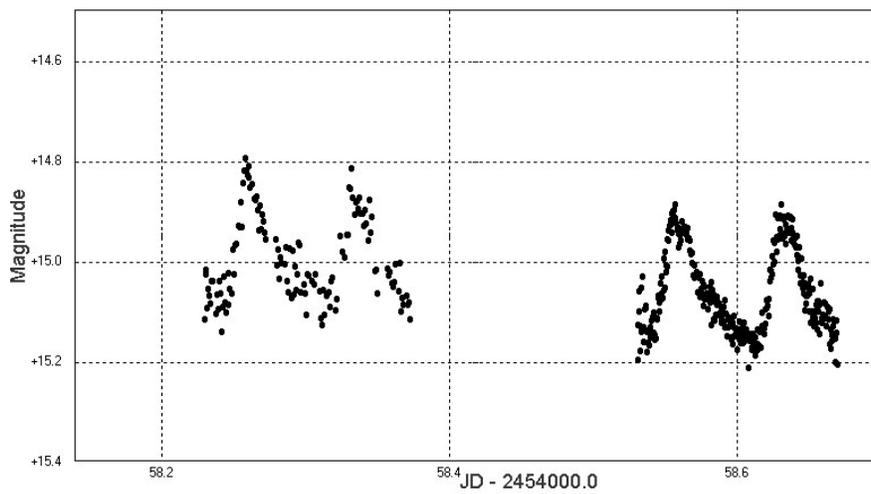
Figure 5: Time-series data from Nov 18 and 19 (R. Pickard and T. Krajci)

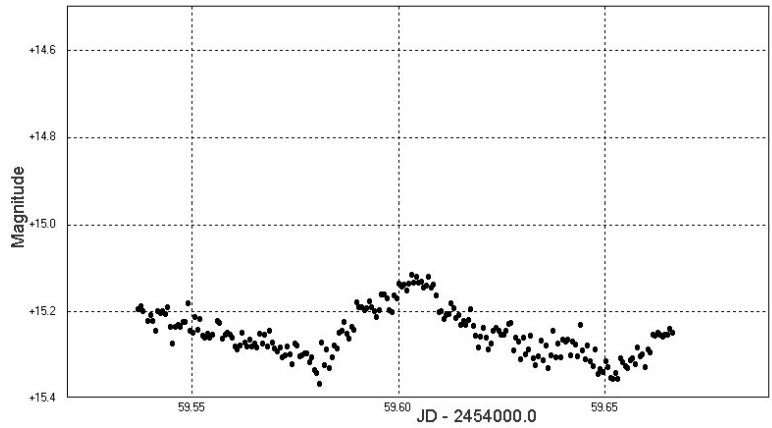
Figure 6: Time-series data from Nov 20 (T. Krajci)

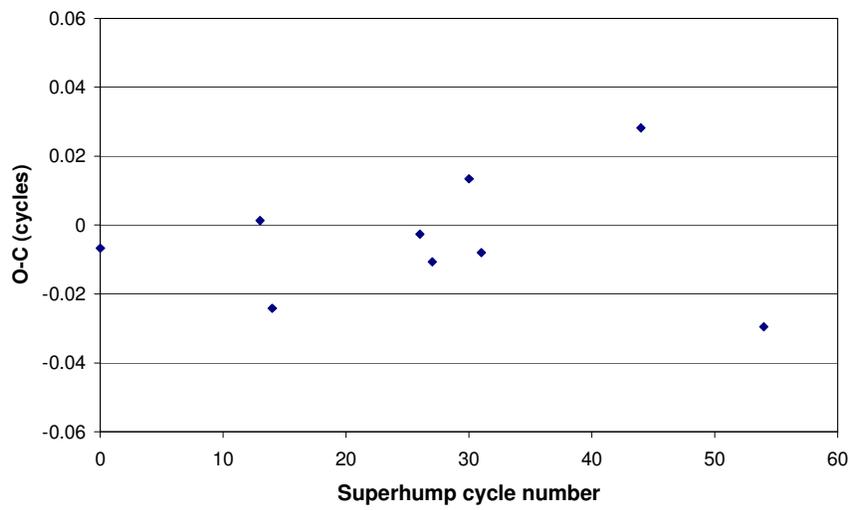
Figure 7: Plot of times of (O-C) residuals versus cycle number

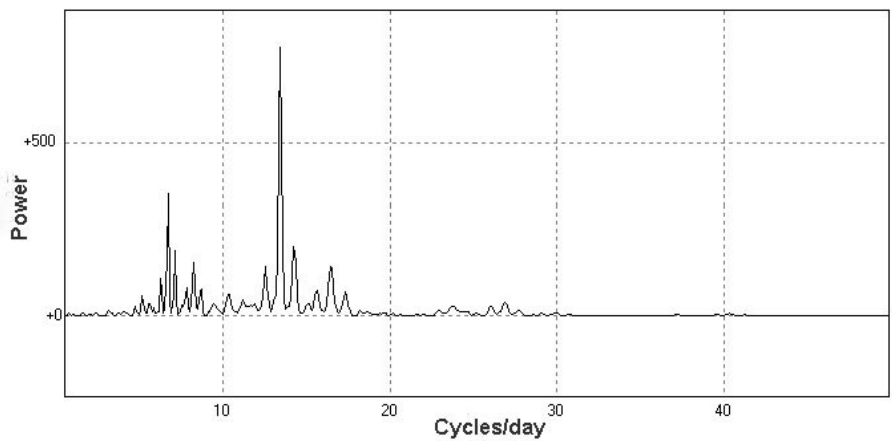
Figure 8: Power spectrum of combined time-series data

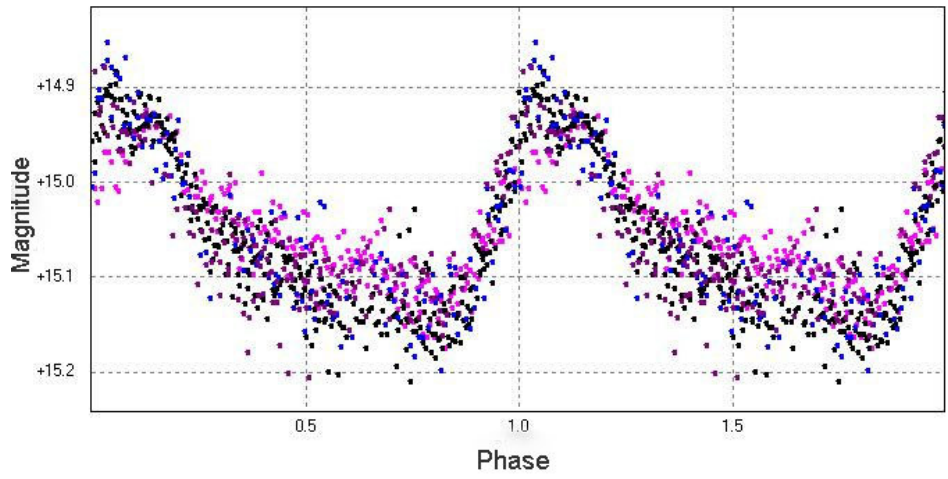

Figure 9: Phase diagram of the time-series data from runs 1 to 6